\documentclass[twocolumn]{aastex631}

\usepackage[version=4]{mhchem}
\usepackage{multirow}
\usepackage{booktabs}

\graphicspath{{./}{figures/}}

\begin{document}

\title{Collisional excitation and non-LTE modelling of interstellar
  chiral propylene oxide}

\author{Karlis Dzenis} \affiliation{University of Edinburgh, School of
  Chemistry, Edinburgh EH9 3FJ, United
  Kingdom}\affiliation{Universit\'e Grenoble Alpes, CNRS, IPAG,
  F-38000 Grenoble, France}\affiliation{Universit\'e Grenoble Alpes,
  D\'epartement de Chimie Mol\'eculaire, F-38000 Grenoble, France}

\author{Alexandre Faure} \affiliation{Universit\'e Grenoble Alpes,
  CNRS, IPAG, F-38000 Grenoble, France}

\author{B.~A. McGuire} \affil{Department of Chemistry, Massachusetts
  Institute of Technology, Cambridge, MA 02139, USA}\affil{National
  Radio Astronomy Observatory, 520 Edgemont Rd., Charlottesville, VA
  22903, USA}

\author{A.~J. Remijan} \affil{National Radio Astronomy Observatory,
  520 Edgemont Rd., Charlottesville, VA 22903, USA}

\author{P.~J. Dagdigian} \affil{Department of Chemistry, The Johns
  Hopkins University, Baltimore, MD 21218-2685, USA}

\author{C. Rist}\affil{Universit\'e Grenoble Alpes, CNRS, IPAG,
  F-38000 Grenoble, France}

\author{R. Dawes} \affil{Department of Chemistry, Missouri University
  of Science and Technology, Rolla, Missouri 65409, USA}

\author{E. Quintas-S\'anchez} \affil{Department of Chemistry, Missouri
  University of Science and Technology, Rolla, Missouri 65409, USA}

\author{F. Lique} \affil{Universit\'e du Rennes, CNRS, IPR (Institut de
  Physique de Rennes) – UMR 6251, F-35000 Rennes, France}

\author{M. Hochlaf} \affil{Universit\'e Gustave Eiffel, COSYS/LISIS, 5
  Bd Descartes, 77454, Champs sur Marne, France}


\begin{abstract}

The first set of theoretical cross sections for propylene oxide
(\ce{CH3CHCH2O}) colliding with cold He atoms has been obtained at the
full quantum level using a high-accuracy potential energy surface. By
scaling the collision reduced mass, rotational rate coefficients for
collisions with para-\ce{H2} are deduced in the temperature range
5$-$30~K. These collisional coefficients are combined with radiative
data in a non-LTE radiative transfer model in order to reproduce
observations of propylene oxide made towards the Sagittarius~B2(N)
molecular cloud with the Green Bank and Parkes radio telescopes. The
three detected absorption lines are found to probe the cold ($\sim
10$~K) and translucent ($n_{\rm H} \sim 2000$~cm$^{-3}$) gas in the
outer edges of the extended Sgr~B2(N) envelope. The derived column
density for propylene oxide is $N_{tot} \sim 3\times
10^{12}$~cm$^{-2}$, corresponding to a fractional abundance relative
to total hydrogen of $\sim 2.5\times 10^{-11}$. The present results
are expected to help our understanding of the chemistry of propylene
oxide, including a potential enantiomeric excess, in the cold
interstellar medium.

\end{abstract}

\keywords{astrochemistry --- molecular data --- molecular processes
  --- ISM: molecules}

\section{Introduction} \label{sec:intro}

Molecular chirality was discovered in 1848 by the French chemist Louis
Pasteur but the term chirality was only introduced in 1894 by Lord
Kelvin \citep{Gal2011}. It was defined by Vladimir Prelog in his 1975
Nobel
lecture\footnote{https://www.nobelprize.org/uploads/2018/06/prelog-lecture.pdf}:
``An object is chiral if it cannot be brought into congruence with its
mirror image by translation and rotation''. Chiral molecules are
components of essential building blocks of all life on Earth,
specifically nucleic acids and proteins. They can exist as
dextrorotatory (D) or levorotatory (L) enantiomers depending on the
way they rotate plane-polarized light, clockwise or counter clockwise,
respectively. Interestingly, life on Earth is exclusively homochiral
because nucleic acids contain only D-sugars while proteins are built
from L-amino acid monomers. While homochirality is seen as a
requirement for the evolution of life \citep{Joyce1987}, its origins
are still unclear and are actively discussed. Several studies suggest
that the origin of a primordial enantiomeric excess ($ee$) is in fact
extra-terrestrial \citep{Bailey1998}. This is consistent with evidence
from laboratory measurements of $ee$ of the L enantiomer of several
amino acids that were found in Murchison and Orgueil meteorite samples
which are the oldest samples with a measured $ee$ \citep{Glavin2009}.

A number of astronomically relevant mechanisms that could produce a
primordial $ee$ have been proposed in the literature. A small
imbalance in chiral symmetry has been demonstrated experimentally to
arise in gas-phase by radiolysis with beta-decay electrons
\citep{Dreiling2014}. Alternatively, homochirality can be produced
through asymmetric photochemistry of molecules by circularly polarised
light (CPL) \citep{Modica2014}. Overall, these clues suggest a
scenario where an enantiomeric enrichment of chiral organic molecules
was produced in molecular clouds of the interstellar medium (ISM)
which in subsequent solar life cycle stages was incorporated in the
formation of a circumstellar disk, new planets, meteorites or
comets. Eventually the (amplified) enantiomerically enriched organic
material would have been delivered to Earth and influenced the
evolution of prebiotic organic molecules.

The prospect of this scenario has motivated further investigations
aimed at detecting signatures of chiral molecules
outside the Earth and the Solar system. To date, more than 250
molecules have been detected in the ISM, about a third of which are
complex organic molecules (COMs) \citep{Herbst2009}, yet only one,
propylene oxide (\ce{CH3CHCH2O}), is a chiral molecule. Propylene
oxide was detected in 2016 towards the centre of the Milky Way galaxy,
in cold, presumably low-density gas of the high-mass star-forming
region Sagittarius~B2(N) \citep{McGuire2016}. With these observations
it was not possible to distinguish between the enantiomers of
propylene oxide and determine if an $ee$ exists. Recently, however, a
Gas Chromatographic – Mass Spectroscopic (GC-MS) analysis of ethanol
extracts of Murchison meteorite samples reported the presence of two
chiral derivatives of propylene oxide with an $ee$
averaging to $\sim 10$\% \citep{Pizzarello2018}. The measured $ee$
from meteorite samples cannot be related to the composition of
propylene oxide detected in the ISM from this study alone, therefore
two questions remain unanswered. Does an $ee$ of propylene oxide exist
in the ISM? If it does, how was it produced? To determine if an $ee$
of propylene oxide can exist in the ISM, it is important to understand
formation pathways to this molecule, which could then be evaluated in
astrochemical models if the abundance is known. Recent experimental
works have shown that propylene oxide can be synthesized in
interstellar ice analogs under irradiation with MeV protons
\citep{Hudson2017} or keV electrons \citep{Bergantini2018}. But other
non-energetic mechanisms, including gas-phase synthesis
\citep{Bodo2019}, may exist.

To elucidate the possible formation pathways, however, it is first
crucial to determine an accurate column density and abundance of
propylene oxide, which constitutes the main aim of this study. Initial
estimates of propylene oxide abundance towards the envelope of Sgr~B2
have been made by \cite{Cunningham2007} who had determined an upper
limit for propylene oxide column density in the Sgr~B2 Large Molecule
Heimat (N-LMH) compact source at $6.7 \times 10^{14}$~cm$^{-2}$ with a
high excitation temperature, $T_{ex} = 200$~K. \cite{McGuire2016}
identified propylene oxide by three rotational transitions in
absorption towards Sgr~B2(N) and their observations were fitted the
best for an excitation temperature $T_{ex}$ = 5.2~K and a column
density of $1 \times 10^{13}$~cm$^{-2}$, which was consistent with the
previously set upper boundary. In the analysis of \cite{McGuire2016}
it was noted that the excitation temperature that fits their
observations best was not well-constrained and other column densities
for an excitation temperature range 5~K $\leq T_{ex} \leq$ 35~K could
also produce a good fit. This was highlighted by some of us
\citep{Faure2019} as a source of uncertainty in the derived
abundance. Additionally, the rotational level populations of propylene
oxide cannot be accurately described by a single excitation
temperature, as it relies on the local thermodynamic equilibrium (LTE)
assumption in which the detected levels follow a Boltzmann
distribution. In general, LTE conditions are not fulfilled due to the
low density of ISM environments, including the envelope of Sgr~B2(N)
\citep{Faure2014}. Instead, a non-LTE approach is necessary to provide
accurate column densities and also to extract the local physical
conditions (density and kinetic temperature).

In 2019, some of us generated the 3D potential energy surface (PES) of
the \ce{CH3CHCH2O}$-$He interacting system \citep{Faure2019}. The
first objective of this study is to use this PES to calculate
rotational cross sections for the inelastic collision between
propylene oxide and He at low collision energy and at the quantum
level. Using standard reduced-mass scaling, rate coefficients for
propylene oxide interacting with para-\ce{H2}, the most abundant
collider in the cold ISM, are then deduced for kinetic temperatures
below 30~K. The second objective is to combine collisional and
radiative rates in a non-LTE model in order to reproduce the detected
rotational lines of propylene oxide by \cite{McGuire2016} and to
determine both an accurate column density for propylene oxide and the
physical conditions in the absorbing cold shell towards Sgr~B2(N).

Section~2 is dedicated to rotational cross section calculations, which
are used to determine rotational (de)excitation rate
coefficients. Radiative transfer calculations are described in
Section~3. Conclusions and future work are discussed in Section~4.

\section{Collisional cross sections and rate coefficients} \label{sec:xsec}

Quantum scattering calculations are based on a high-accuracy
\ce{CH3CHCH2O}$-$\ce{He} PES which was previously computed,
characterised and presented by \citet{Faure2019}. Briefly, this PES
was determined using the explicitly correlated coupled-cluster theory
extended to the basis set limit [CCSD(T)-F12b/CBS] and it was
accurately fitted to spherical harmonics. The global minimum of the
fitted PES was found at -79.5~cm$^{-1}$ relative to the energy of
separated molecules. We note that this value is in very good agreement
(within 1~cm$^{-1}$) with the global minimum of the
\ce{CH3CHCH2O}$-$\ce{He} PES built from the analysis of experimental
scattering data \citep{Palazzetti21}.

In the study of \citet{Faure2019}, rotational cross sections were
computed for a single collision energy of 10~cm$^{-1}$. Here we have
extended the calculations to cover an extensive grid of total energies
(0.5$-$213~cm$^{-1}$) and then derive rate coefficients in the
temperature range 5$-$30~K. Because we are only concerned with energy
transfer between helium and randomly oriented propylene oxide, chiral
effects are entirely ignored.

Cross section calculations presented below employed three
assumptions. First, it was assumed that cross sections for propylene
oxide colliding with para-H\textsubscript{2} are equal to those with
He. This assumption should be appropriate for molecules with more than
three heavy atoms if the collision partner is para-H\textsubscript{2}
in its ground rotational state ($j$=0) as found for collisional
excitation of HC\textsubscript{3}N by He, ortho-H\textsubscript{2} and
para-H\textsubscript{2} \citep{Wernli2007,Wernli2007b}. The population
of ortho-H\textsubscript{2} can be considered insignificant compared
to para-H\textsubscript{2} below 30~K \citep{Faure2013}. Second,
propylene oxide was treated as a rigid rotor. This approximation is
justified for scattering calculations at low collision energy, as
demonstrated by \citet{Faure2016} for the CO$-$H$_2$ system. Third,
due to internal rotation (methyl torsion), rotational levels of
propylene oxide are split by tunneling into doublets, the
nondegenerate $A$ levels and the doubly degenerate $E$ levels
\citep{Herschbach1958}. In the ground torsional state, however, the
splitting could not be resolved in the astronomical observations
because the barrier to internal rotation is very high
\citep{Stahl2021}. As a result, we used a single rigid-rotor
asymmetric top hamiltonian with experimental ground-state rotational
constants $A=0.6012$~cm$^{-1}$, $B$=0.2229~cm$^{-1}$ and
$C$=0.1985~cm$^{-1}$ taken from the Cologne Database for Molecular
Spectroscopy (CDMS) \citep{Muller2005}.

Cross section calculations were performed using the \texttt{Hibridon}
code \citep{hibridon}, which is based on solving the time-independent
close-coupling (CC) equations. The general theory for quantum
scattering calculations within the CC formalism, introduced by
\citet{Garrison1976} for an asymmetric rotor colliding with a
structureless atom, was extended to molecules like propylene oxide
that do not possess any symmetry elements by \citet{Faure2019} and was
applied here. The second-order differential CC equations were
integrated using the hybrid propagator of \citet{Alexander1987} with
propagation parameters as in \citet{Faure2019}. Cross sections for
purely rotational transitions were calculated for a total energy range
below the first torsional mode of propylene oxide at 213~cm$^{-1}$
\citep{Sebestik2011}, with an energy step of 0.5~cm$^{-1}$ below
$E_{tot}=80$~cm$^{-1}$ to account for the numerous resonances arising
from quantum effects. The rotational constants of propylene oxide, as
listed above, are moderately small and the rotational spectrum
relatively dense, therefore solving CC equations was expected to be
computationally demanding with respect to CPU time and memory.

The rotational states are defined as $j_{K_a K_c}$ where $j$ is the
angular momentum quantum number, and $K_a$ and $K_c$ are projections
of $j$ along the $a$ and $c$ axes, which by convention are the axes
with the smallest and largest moments of inertia, respectively. A
basis of internal states with $j_{max}$=22 was selected to ensure that
inelastic cross sections for all transitions among the lowest 100
rotational levels (up to $11_{29}$ at 29.862~cm$^{-1}$)
were converged to within 20\%.
As discussed in \citet{Faure2019}, the convergence of cross section
calculations was found to be slow due to couplings to many
energetically inaccessible rotational states (closed channels). To
limit the computational expenses while maintaining accuracy, two
additional parameters, threshold rotational energy ($E_{max}$) and
total angular momentum ($J$), were optimised in the total energy range
to reach a global convergence of total cross sections (summed over
$J$) better than 20\%. This was achieved with
$E_{max}$=160$-$213~cm$^{-1}$ and for partial waves
$J\leq 50$. The largest calculations involved more than 7300
coupled-channels and about 30 CPU hours for one partial wave on Intel
Gold processors. A total of $\sim 2\times 10^5$ CPU hours were
consumed for the whole project.

\begin{figure}[ht!]
\includegraphics*[scale=0.38,angle=0.]{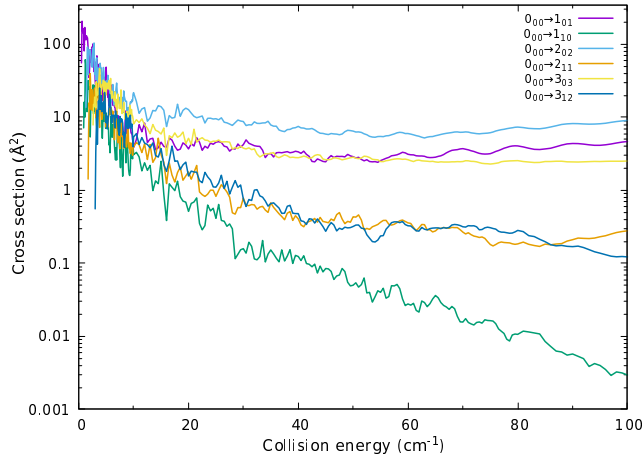}
\caption{Integral cross sections for state-to-state rotational
  excitation of propylene oxide from the rotational ground state
  ($0_{00}$) by collisions with He.}
\label{fig:figure1}
\end{figure}

Integral cross sections for excitations from the ground state to
rotational levels that are relevant to the observed absorption lines
({\it i.e.}, $1_{01}$, $1_{10}$, $2_{02}$, $2_{11}$, $3_{03}$ and
$3_{12}$) are presented as a function of collision energy ($E_{col}$)
in Figure~\ref{fig:figure1}.
In general, the calculated cross sections are highest at low collision
energies with many resonances that contribute to the peak-trough
pattern and decrease with increasing collision energies until reaching
a plateau. As expected, the cross sections at
10~cm$^{-1}$ are in excellent agreement with those of
\citet{Faure2019}.

\begin{figure}[ht!]
\includegraphics*[scale=0.38,angle=0.]{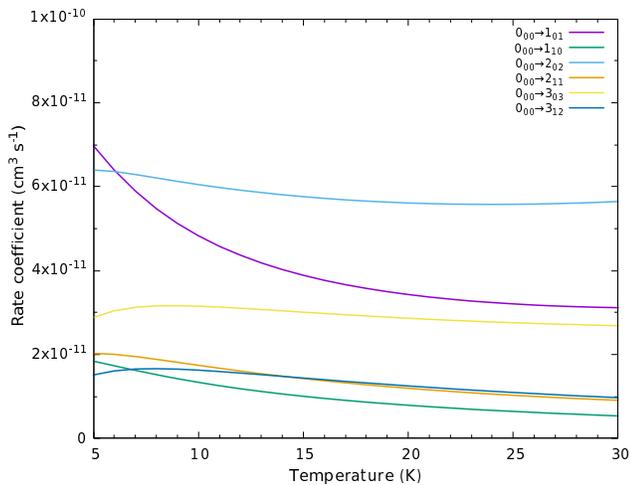}
\caption{Rate coefficients for rotational excitation of propylene
  oxide from the rotational ground state ($0_{00}$) by
  para-H\textsubscript{2}.}
\label{fig:figure2}
\end{figure}

Rate coefficients $k(T)$ for each collisional transition ($j_{K_aK_c}
\to j'_{K'_aK'_c}$) were calculated in the temperature range
$T_{kin}=$5$-$30~K by integrating the total cross sections ($\sigma$)
over a Maxwell--Boltzmann distribution of velocities, which are
directly related to the collision (kinetic) energy ($E_{col}$)
according to Equation~(1):
\begin{equation}
k(T_{kin})=\sqrt{\frac{8k_BT_{kin}}{\pi\mu}} \int_0^{\infty} \sigma(x)
x\exp(-x)dx,
\label{equation1}
\end{equation}
where $x=E_{col}/(k_BT_{kin})$, $k_B$ is the Boltzmann constant and
$\mu$ is the collisional reduced mass. Using the aforementioned
assumption, that cross sections for collisions with He are equal to
those with para-H\textsubscript{2}, the reduced mass of propylene
oxide and H\textsubscript{2} ($\mu$=1.94800083~amu) was used. Rate
coefficients are displayed as a function of temperature for excitation
from the ground state to rotational levels relevant to the observed
absorption lines (\textit{i.e.}, $1_{01}$, $1_{10}$, $2_{02}$,
$2_{11}$, $3_{03}$ and $3_{12}$) in Figure~\ref{fig:figure2}.
Rate coefficients clearly follow the ordering of cross sections in
Figure~\ref{fig:figure1} due to their relation through
Eq.~(\ref{equation1}), however their dependence on temperature is
smooth compared to the pattern of cross sections which is a result of
the Maxwell--Boltzmann averaging. The favoured transition with the
largest rate coefficient is 0\textsubscript{00} $\to$
2\textsubscript{02} which corresponds to a propensity rule $\Delta
j$=2 and $\Delta K_a$=0 while the next two highest rate coefficients
are observed for transitions $0_{00} \rightarrow 1_{01}$ and
$0_{00}\rightarrow 3_{03}$, respectively. This trend indicates the
conservation of the angular momentum along the axis of smallest moment
of inertia ($a$ axis), since $\Delta K_a$=0. Interestingly, such
ordering of collisional rate coefficients for transitions from the
ground state corroborates the detection of rotational transitions of
propylene oxide (\textit{i.e.}, $1_{01} \rightarrow 1_{10}$,
$2_{02}\rightarrow 2_{11}$ and $3_{03} \rightarrow 3_{12}$) in
absorption. Nevertheless, the real populations of rotational levels
can only be determined by considering also the radiative processes in
a non-LTE model for specific physical conditions.

\section{Non-LTE modelling} \label{sec:model}

The non-LTE radiative transfer calculations were performed with
\texttt{Radex} \citep{VanderTak2007} using the escape probability
approximation for a uniform sphere. The model incorporated both the
calculated collisional rate coefficients among 100 lowest rotational
levels (a total of 4950 de-excitation transitions) and Einstein
coefficients for radiative transitions obtained from the CDMS database
\citep{Muller2005}. The only ambient radiation field assumed is the
cosmic microwave background (CMB) and we omitted, in particular, the
diluted infrared emission from the hot cores Sgr~B2(M) and Sgr~B2(N),
as well as the extended microwave continuum from the Galactic Center
region. These neglected sources of radiative pumping (absorption and
stimulated emission) can possibly compete with collisional processes
and they will be investigated in future works. The \texttt{Radex} code
was applied to calculate excitation temperature ($T_{ex}$), and
opacity ($\tau$) of each transition at physical conditions specified
by a kinetic temperature ($T_{kin}$), density of H\textsubscript{2}
($n({\rm H_2})$) and column density of propylene oxide ($N_{tot}$). A
range for each physical parameter was selected to compute a grid of
non-LTE models, in particular $T_{kin}$ was constrained to 5$-$30~K,
$n({\rm H_2})$ was restricted to 10$-$10$^6$~cm$^{-3}$ and $N_{tot}$
to $10^{12}-10^{14}$~cm$^{-2}$. A linewidth, full width at half
maximum (FWHM), of 13~km~s$^{-1}$ was used as determined by
\citet{McGuire2016}.

For each transition a measurable line intensity at a frequency $\nu$,
(corrected) antenna temperature $T_a^*$ (in Kelvin), can be expressed
in terms of $T_{ex}$ and $\tau$ as follows:

\begin{equation}
    T_a^*(\nu)=(\eta_bBJ_{\nu}(T_{ex} )-\eta_bBJ_{\nu}(T_{cmb} )-T_c
    (\nu))(1-\exp(-\tau)),
    \label{equation2}
\end{equation}
where the Green Bank Telescope (GBT) beam efficiency $\eta_b$ and the
main beam continuum temperature
$T_c(\nu)=10^{-1.06\log_{10}(\nu[GHz])+2.3}$ were taken from
\citet{Hollis2007}, $J_{\nu}(T)$=$(h\nu/k_B)/(e^{h\nu/k_BT}-1)$,
$T_{cmb}$ is the CMB temperature (2.73~K), and $B$ is the GBT beam
dilution factor accounting for the spatial overlap of the Sgr~B2(N)
continuum with the GBT beam (both assumed to have a gaussian shape),
$B=\theta_s^2/(\theta_s^2+\theta_b^2)$, with $\theta_s$ and $\theta_b$
as the source and beam sizes, respectively. The size of the background
continuum source against which propylene oxide absorbs was taken as
$\theta_s = 20''$, as in \citet{McGuire2016}. We note that this value
is consistent with $\theta_s = 77''(\nu)^{-0.52}$, as derived by
\citet{Hollis2007} (for the north component) when $\nu \sim$
12--14~GHz. The GBT half-power beamwidth was taken as $\theta_b=
754''/\nu$, where $\nu$ is in units of GHz.

Importantly, only transitions $2_{02}\rightarrow 2_{11}$ and
$3_{03}\rightarrow 3_{12}$ were detected by GBT, while transition
$1_{01}\rightarrow 1_{10}$ was detected with a beam almost twice the
size of GBT by the Parkes Radio Telescope and incorporated a
contribution to the continuum background from an adjacent source,
Sgr~B2(M). Therefore, in our quantitative model we focus on the two
lines observed by GBT, namely $2_{02}\rightarrow 2_{11}$ and
$3_{03}\rightarrow 3_{12}$ and we neglect the contribution of sources
other than Sgr~B2(N). Thus, the observed intensity of the Parkes line
cannot be compared to our predictions.

\begin{figure}[ht!]
\includegraphics*[scale=0.38,angle=0.]{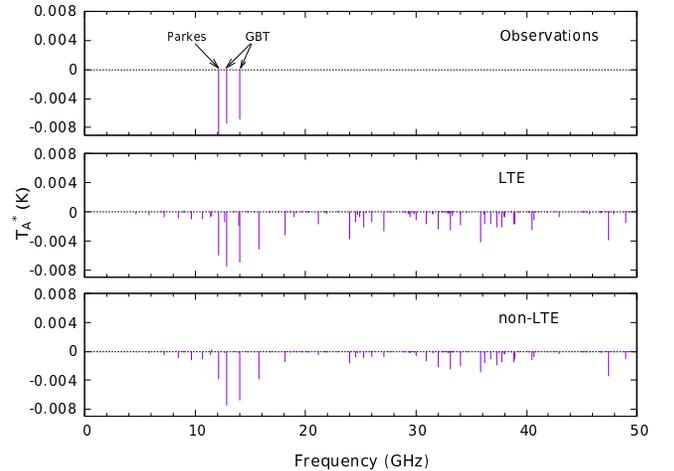}
\caption{Observed (top), modelled LTE (middle) and non-LTE (bottom)
  line intensities ($T_a^*$) of rotational transitions of propylene
  oxide in the frequency range of 1$-$50~GHz.}
\label{fig:figure3}
\end{figure}

Physical parameters that best model the observed $T_a^*$ of the two
absorption lines detected by GBT were determined using a least-squares
fit. The Parkes line was also employed and restricted to be in
absorption, {\it i.e.} with negative $T_a^*$ values. The best fit
non-LTE model is compared to the two observed GBT absorption line
intensities in Figure~\ref{fig:figure3}. We note that a local standard
of rest velocity ($V_{LSR}$) of +64~km.s$^{-1}$, characteristic of the
extended envelope of Sgr~B2(N), was assumed for the detected
lines. Other details about GBT and Parkes observations can be found in
\citet{McGuire2016}. Additionally, we present in the middle panel of
Figure~\ref{fig:figure3} a LTE model where all lines are constrained
to have the same excitation temperature.

In our LTE model, the density was taken as $n({\rm H_2})=1\times
10^{10}$~cm$^{-3}$ in order to guarantee LTE
populations. Indeed, the critical densities\footnote{The critical
  density for a particular level is given by the ratio between the sum
  of all radiative rates and the sum of all deexcitation collisional
  rates from this level.} for the detected levels $1_{01}$, $2_{02}$
and $3_{03}$ are $\sim 10^3$~cm$^{-3}$. The excitation temperature
$T_{ex}=T_{kin}=5.2$~K and column density $N_{tot}=9\times
10^{12}$~cm$^{-2}$ predicts the observed GBT line
intensities closely (within 0.1~mK) which aligns with LTE modelling
results by \citet{McGuire2016} who found the best fit with the same
$T_{ex}$ and $N_{tot}=1\times 10^{13}$~cm$^{-2}$.

The non-LTE calculations provide the best fit to the GBT line
intensities for model parameters $n({\rm H_2})=1000$~cm$^{-3}$,
$T_{kin}$=8~K and $N_{tot}=3\times 10^{12}$~cm$^{-2}$. The resulting
root-mean-square (rms) error is very low ($< 0.1$~mK), as expected for
an overfit model. Exploring the 3D grid of $\chi^2$ values shows that
the kinetic temperature and \ce{H2} density are poorly constrained
while the best-fit value for the propylene oxide column density
corresponds to a steep minimum. It should be noted that our best-fit
model is also consistent with the non-detection of other rotational
lines above the continuum noise by previous surveys in the range
0--120~GHz \citep{Cunningham2007,McGuire2016}. In particular, we
report in Table~1 the non-LTE predicted intensity for the strongest
transitions ($>$1.5~mK) in the range 1--50~GHz alongside their
opacity, excitation temperature and the noise level of the PRIMOS
observations or the cause of a non-detection (missing data, radio
frequency interference (RFI))\footnote{Data are available from the
  NRAO archive at https://archive.nrao.edu under GBT project code
  AGBT07A$\_$051.}.

\begin{table}
\centering
\caption{Opacity, excitation temperature and intensity of the (13)
  strongest absorption lines in our best-fit non-LTE model below 50
  GHz, along with the noise level of the PRIMOS observations.}
\begin{tabular}{ccccc} \toprule
{$\nu$ (GHz)} & {$\tau$} & {$T_{ex}$ (K)} & {$T^{*}_{a}$ (K)} & {Noise level (K)} \\ \midrule
12.0724	& 2.734$\times10^{-4}$ & 4.09 & -0.0039 & RFI\\
12.8373 & 5.580$\times10^{-4}$ & 2.74 & -0.0074 & Detected Feature\\
14.0478 & 5.556$\times10^{-4}$ & 2.61 & -0.0067 & Detected Feature\\
15.7751 & 3.542$\times10^{-4}$ & 2.89 & -0.0038 & No Data\\ 
18.1036 & 1.651$\times10^{-4}$ & 3.38 & -0.0015 & 0.0035\\
23.9752 & 2.586$\times10^{-4}$ & 5.81 & -0.0016 & 0.0053\\
32.0418 & 3.965$\times10^{-4}$ & 2.15 & -0.0021 & 0.0229\\
33.1211 & 5.021$\times10^{-4}$ & 2.25 & -0.0025 & 0.0075\\
34.0591 & 4.055$\times10^{-4}$ & 2.45 & -0.0020 & 0.0075\\
35.8779 & 7.083$\times10^{-4}$ & 4.11 & -0.0029 & 0.0074\\
36.2171 & 3.639$\times10^{-4}$ & 2.65 & -0.0016 & 0.0100\\
37.3340 & 4.361$\times10^{-4}$ & 2.50 & -0.0019 & 0.0104\\
47.4285 & 1.105$\times10^{-3}$ & 3.52 & -0.0034 & 0.0085\\
    \bottomrule
\end{tabular}
\end{table}
  
In Figure~\ref{fig:figure3} both LTE and non-LTE spectra display a
potential signature of propylene oxide at 15.8~GHz in absorption with
similar intensities of $-5.5$~mK and $-3.8$~mK, respectively. This
line is $\sim 1.8$ times weaker than the faintest detected transition
$3_{03}\rightarrow 3_{12}$ at 14.0~GHz and corresponds to the
transition $4_{04}\rightarrow 4_{13}$ which was not detected with the
GBT or Parkes telescope due to absence of a corresponding receiver
\citep{McGuire2016}. More generally, we note that although the derived
column density is a factor of 3 larger, the LTE spectrum agrees
surprisingly well with the non-LTE spectrum. This can be explained by
similar opacities (in the range $2.7-5.8\times 10^{-4}$), and
excitation temperatures (in the range $2.6-5.2$~K) for the three
detected lines, which both contribute to $T_a^*$. Still, some lines
are predicted to be inverted ({\it i.e.}  with negative $T_{ex}$) with
the non-LTE model, although none of these potential weak masers have
sufficient opacities to become detectable in this source.

The uncertainty in the three model parameters ($T_{kin}$, $n({\rm
  H_2})$ and $N_{tot}$) is difficult to assess statistically with only
three detected lines. Assuming the main source of uncertainty is
introduced by the measurements as reported by \citet{McGuire2016}, we
have considered the 1$\sigma$ deviation limits for the intensities of
transitions $2_{02}\rightarrow 2_{11}$ ($\pm$5~mK) and
$3_{03}\rightarrow 3_{12}$ ($\pm$6~mK) and the corresponding interval
in the three physical parameters. The best-fit models at the 1$\sigma$
limits of line intensities constrain the kinetic temperature to the
range $T_{kin}=$5--13~K, the H$_2$ density to the range $n({\rm
  H_2})=$800--1900~cm$^{-3}$ and the column density to the range
$N_{tot}=3\times 10^{12}$--$4\times 10^{12}$~cm$^{-2}$. We cannot,
however, exclude kinetic temperatures larger than 13~K: $T_{kin}$ in
the range 20--30~K also provide reasonable results, but with larger
rms error.

\begin{figure}[ht!]
\includegraphics*[scale=0.38,angle=0.]{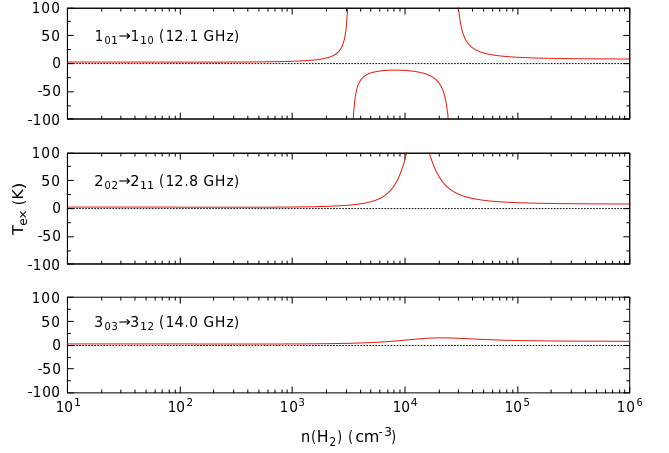}
\caption{Excitation temperature as a function of H\textsubscript{2}
  volume density ($n({\rm H_2})$) for the observed rotational
  transitions of propylene oxide towards Sgr~B2~(N), $1_{01} \to
  1_{10}$ (12.1~GHz), $2_{02} \to 2_{11}$ (12.8~GHz) and $3_{03} \to
  3_{12}$ (14.0~GHz) at $T_{kin}$=8~K and $N_{tot}=3\times
  10^{12}$~cm$^{-2}$.}
\label{fig:figure4}
\end{figure}

\begin{figure}[ht!]
\includegraphics*[scale=0.38,angle=0.]{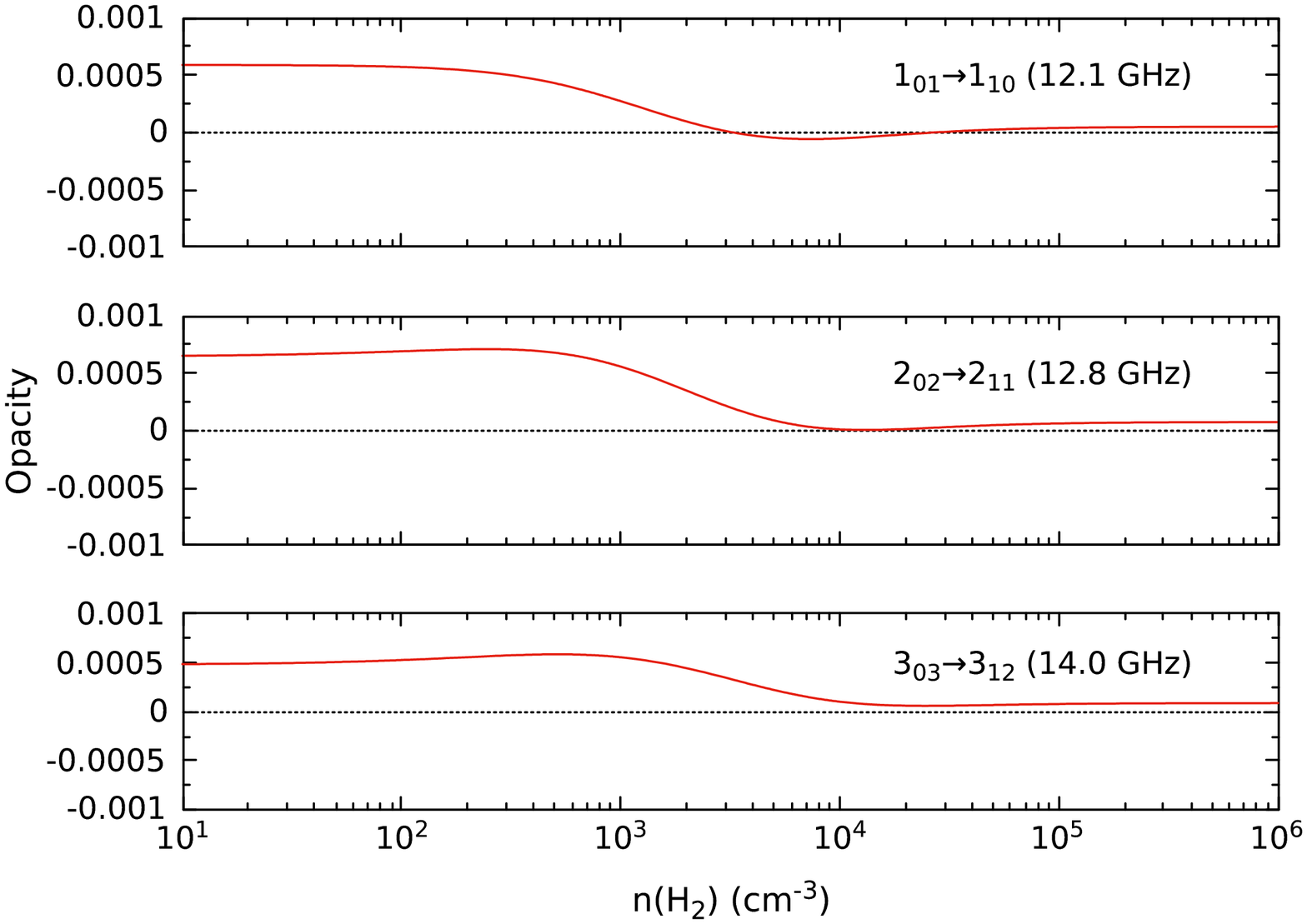}
\caption{Opacity as a function of H\textsubscript{2} volume density
  ($n({\rm H_2})$) for the observed rotational transitions of
  propylene oxide towards Sgr~B2~(N), $1_{01} \to 1_{10}$ (12.1 GHz),
  $2_{02} \to 2_{11}$ (12.8 GHz) and $3_{03} \to 3_{12}$ (14.0 GHz) at
  $T_{kin}$=8~K and $N_{tot}=3\times10^{12}$~cm$^{-2}$.}
\label{fig:figure5}
\end{figure}

It is useful to analyse the excitation temperatures $T_{ex}$ and
opacities $\tau$ of all three detected transitions. $T_{ex}$ and
$\tau$ are displayed as a function of $n({\rm H_2})$ in
Figures~\ref{fig:figure4} and \ref{fig:figure5} for the best-fit
non-LTE model with parameters $T_{kin}$=8~K and $N_{tot}=3\times
10^{12}$~cm$^{-2}$.
For all three transitions, $T_{ex}$ and $\tau$ are positive, except in
the range $3\times10^{3}$~cm$^{-3}$ $\lesssim n({\rm
  H_2}) \lesssim 3\times 10^{4}$~cm$^{-3}$ where the
Parkes line at 12.1~GHz gets inverted and the molecule is capable of
masing (opacity is however very low). When $n({\rm H_2})\gtrsim
3\times 10^{4}$~cm$^{-3}$, all opacities are positive and
small, while $T_{ex}$ gets closer to $T_{kin}$ as the system
approaches LTE. Since all three lines were detected in absorption,
only the range of $n({\rm H_2})$ where $T_{ex}$ and $\tau$ are
positive for all transitions can be considered. The opacity further
constrains the \ce{H2} volume density, as in the range $n({\rm H_2})>
3\times 10^{4}$~cm$^{-3}$ the opacities are very low
($\sim 1\times 10^{-5}$) and the absorption lines would not be
detected above the continuum noise. As a result, the non-LTE
calculations strongly restrict the observations of propylene oxide to
a low-density region in the envelope of Sgr~B2(N) with $n({\rm H_2})<
3\times 10^{3}$~cm$^{-3}$. At higher densities, opacities
become so low that the lines would be undetectable.

Finally, we note that a non-LTE model for propylene oxide was
presented recently by \citet{Das2019}. In their work, the authors
employed the collisional rate coefficients of \ce{CH3OH} with \ce{H2},
which is obviously a crude approximation to mimic the
\ce{CH3CHCH2O}$-$\ce{H2} system. Their best model was found for
$T_{kin}=$5--10~K and $n({\rm H_2})>10^4$~cm$^{-3}$. The high
densities found by these authors likely reflect the approximate
collisional data, as well as the propylene oxide column density kept
constant at $1\times 10^{13}$~cm$^{-2}$.

\section{Discussion} \label{sec:discussion}

The presented non-LTE calculations suggest that propylene oxide was
detected towards a cold, $T_{kin}\sim 10$~K, and moderately dense
$n({\rm H_2}) \sim 1000$~cm$^{-3}$ region towards Sgr~B2(N) with a
column density $N_{tot}\sim 3\times 10^{12}$~cm$^{-2}$. The determined
column density of propylene oxide was found to be robust and it is
also in relative good agreement with previous estimates: it is a
factor of $\sim$3 lower than the LTE value estimated by
\citet{McGuire2016} and, as expected, is below the upper limit at
$6.7\times 10^{14}$~cm$^{-2}$ set by \citet{Cunningham2007}. The
density and in particular the kinetic temperature are less well
determined. They are still consistent with detection of other COMs
towards the extended cold envelope of Sgr~B2(N), with low and
sub-thermal rotational temperatures ($<10$~K) (see \citet{Corby2015}
and references therein).

The derived physical parameters are characteristic of {\it
  translucent} regions, which are not commonly associated with the
presence or chemistry of COMs due to the presence of the (only
partially shielded) UV interstellar radiation field that contributes
to dissociation of large molecules. The formation of COMs is typically
thought to occur in more dense and shielded regions, like cold
prestellar cores and hot cores, where the proton density ($n_{\rm
  H}=2n({\rm H_2})$) is larger than $\sim 10^5$~cm$^{-3}$
\citep{Bacmann2012}. But COMs have been also detected in UV-irradiated
photon dominated regions (PDR) \citep{Gratier2013}. Towards the
envelope of Sgr~B2(N), the series of \ce{HCOOCH3} weak maser lines
below 30~GHz were found to be associated with lower densities $n_{\rm
  H}\sim 3\times 10^4$~cm$^{-3}$ \citep{Faure2014}. Here the derived
density is an order of magnitude lower and may be associated with the
outer edges of the extended Sgr~B2(N) envelope. Our results therefore
question the chemistry and formation of complex molecules in
translucent conditions. This is not a unique result, as recent
analysis of (GBT) PRIMOS and (ALMA) EMoCA survey data also identified
several COMs in translucent clouds along the line of sight to
Sgr~B2(N) \citep{Corby2016,Thiel2017,Thiel2019}. We note in this
context the very recent work by \cite{Wang21} on the chemistry of
complex organic molecules in the extended region of Sgr~B2, where the
authors emphasize the possible role of X-ray flares on the reactive
desorption process.


Indeed, the detection of gas-phase COMs at temperatures well below the
thermal desorption temperature has been noted in a variety of
environments in the last several years. These include species like
\ce{CH2OHCHO}, \ce{CH3CHO}, \ce{CH2CHCHO}, \ce{CH3CH2CHO} and
\ce{CNCHO}, all first detected towards Sgr~B2 \citep[see][and
  references therein]{Remijan2008}. \citet{RequenaTorres2006}
conducted an extensive survey of complex molecules in the Galactic
center, noting a number of species with substantially sub-thermal
excitation temperatures. One of these sources, G+0.693-0.027, has seen
substantial recent interest with ALMA observations, resulting in a
number of detections of new organic species, all well-below the
kinetic temperature (see, {\it e.g.}, \citealt{Rivilla2020}). Perhaps
even more striking are the observations of COMs in the cold, dark
starless core TMC-1 (see, {\it e.g.}, \citealt{Agundez2021}), where
the kinetic temperatures are well known to be substantially below the
thermal desorption threshold.  A number of physical and physiochemical
processes have been theorized to account for the non-thermal
desorption of COMs from grain surfaces \citep{Paulive2020}, as well as
substantial quantum chemical efforts suggesting potential new pathways
for gas-phase formation \citep{Balucani2018}.  A generalized picture
of these processes, and under what conditions they dominate, remains
absent.  The detection and quantification of more COMs under
sub-thermal conditions, especially using non-LTE methods that provide
insight into the kinetic temperature and density conditions, is
clearly needed.

Another interesting constraint is provided by the detection of acetone
and propanal, two structural isomers of propylene oxide, also observed
in absorption towards Sgr~B2(N) by \citet{McGuire2016}. Excitation
temperatures of 6.2~K were determined, suggesting similar excitation
conditions, with column densities of $2.1 \times 10^{14}$~cm$^{-2}$
for acetone and $6 \times 10^{13}$~cm$^{-2}$ for propanal
\citep{McGuire2016}. While the relative column densities of the three
\ce{C3H6O} isomers follow thermodynamics, their similar values in such
a cold environment indicates a strongly non-equilibrum and kinetically
controlled chemistry (propylene oxide is less stable than propanal by
about 13,000~K, see \cite{Bergantini2018}).

Finally, we can try to estimate the abundance, relative to the total
hydrogen density, of propylene oxide in the outer shell of the
Sgr~B2(N) envelope. Assuming the low-density layer in front of
Sgr~B2(N), where $n({\rm H_2})=1000$~cm$^{-3}$, extends over $\sim
19$~pc, as described in \cite{Schmiedeke16}, the foreground H$_2$
column density is $\sim 6\times 10^{22}$~cm$^{-2}$ and the fractional
abundance of propylene oxide is $\sim 2.5 \times 10^{-11}$. This is
likely a lower limit since the actual fraction of molecular hydrogen
occupying the same volume as propylene oxide is hard to estimate.


\section{Conclusion} \label{sec:conclusion}

We have presented the first set of collisional rate coefficients for
the rotational excitation of propylene oxide by para-H$_2$ in the
temperature range 5--30~K. The scattering calculations were performed
on the \ce{CH3CHCH2O}--He PES of \cite{Faure2019} with the
\texttt{Hibridon} code. Collisional and radiative Einstein
coefficients were then combined in a non-LTE radiative transfer model
in order to reproduce the three absorption lines of propylene oxide
detected by \citet{McGuire2016}. Our best-fit model suggests that
these transitions sample a cold ($T_{\rm kin}\sim 10$~K), translucent
($n_{\rm H}\sim 2 \times 10^3$~cm$^{-3}$) and extended region
surrounding Sgr~B2(N). Such low densities combined with the relatively
strong and extended continuum background towards Sgr~B2(N) can
sucessfully explain many of the observed absorption spectra of
COMs. On the other hand, weak masers as studied by \citet{Faure2014}
for \ce{HCOOCH3} and \citet{Faure2018} for \ce{CH2NH}, probe a region
with larger density $n_{\rm H}\sim 10^4$~cm$^{-3}$.

We have derived a column density for propylene oxide of $\sim 3\times
10^{12}$~cm$^{-2}$, in good agreement with previous estimates. This
value is a factor of $\sim 10-100$ lower than the column densities
estimated by \citet{McGuire2016} for the more stable isomers propanal
and acetone in the same source. The relative column densities of the
three \ce{C3H6O} isomers should provide another strong constraint on
chemical models, but non-LTE calculations are necessary to model
propanal and acetone observations, which in turn requires the
knowledge of collisional coefficients for these two species.

The constraints and accuracy of the physical parameters predicted by
our non-LTE calculations were, in part, determined by the available
collisional data and the small number of propylene oxide lines
detected in Sgr~B2(N). Our model does not predict additional
detectable lines towards Sgr~B2(N) (except possibly at 15.8~GHz), but
it was not possible to explore kinetic temperatures above 30~K. In
this context, we note the recent experimental work by
\citet{Stahl2021} that should help to assign rotational transitions in
the first excited torsional state $\nu_{24}$ of propylene oxide. Such
transitions could be particularly relevant for the detection of
propylene oxide in warm, $T_{kin}>100$~K, sources. This will require
extension of the present work to include the methyl torsion of
propylene oxide, a great challenge for theory.

Finally, in the absence of high-precision polarization measurements,
it is currently not possible to distinguish enantiomers in the
ISM. The accurate column density and the physical conditions extracted
from our non-LTE model however provides important contraints to
elucidate the formation pathways to propylene oxide in Sgr~B2(N) and
therefore to assess the relevance of the various mechanisms able to
produce a primordial $ee$ in molecular clouds.

\section{Acknowledgements}
This work made use of GBT data from project AGBT07A\_051. The National
Radio Astronomy Observatory is a facility of the National Science
Foundation operated under cooperative agreement by Associated
Universities, Inc. The Green Bank Observatory is a facility of the
National Science Foundation operated under cooperative agreement by
Associated Universities, Inc. This work has been supported by the
French INSU/CNRS Program ``Physique et Chimie du Milieu
Interstellaire'' (PCMI). K.D. acknowledges support by the ERASMUS+
program from European Commission. A.F. and K.D. thank Carlos P\'erez
del Valle and Rafal Szabla for useful discussions. R.D. and
E.Q.-S. are supported by the U.S. Department of Energy (Award
DE-SC0019740). Finally we thank our two referees for their very
constructive comments which have improved the paper.

\bibliography{Dzenis_rev}

\begin{thebibliography}{}
\expandafter\ifx\csname natexlab\endcsname\relax\def\natexlab#1{#1}\fi
\providecommand{\url}[1]{\href{#1}{#1}}
\providecommand{\dodoi}[1]{doi:~\href{http://doi.org/#1}{\nolinkurl{#1}}}
\providecommand{\doeprint}[1]{\href{http://ascl.net/#1}{\nolinkurl{http://ascl.net/#1}}}
\providecommand{\doarXiv}[1]{\href{https://arxiv.org/abs/#1}{\nolinkurl{https://arxiv.org/abs/#1}}}

\bibitem[{{Ag{\'u}ndez} {et~al.}(2021){Ag{\'u}ndez}, {Marcelino}, {Tercero},
  {Cabezas}, {de Vicente}, \& {Cernicharo}}]{Agundez2021}
{Ag{\'u}ndez}, M., {Marcelino}, N., {Tercero}, B., {et~al.} 2021, Astronomy and
  Astrophysics, 649, L4, \dodoi{10.1051/0004-6361/202140978}

\bibitem[{Alexander \& Manolopoulos(1987)}]{Alexander1987}
Alexander, M.~H., \& Manolopoulos, D.~E. 1987, The Journal of Chemical Physics,
  86, 2044, \dodoi{10.1063/1.452154}

\bibitem[{Alexander {et~al.}(2021)Alexander, Manolopoulos, Werner, Follmeg,
  Dagdigian, \& et~al.}]{hibridon}
Alexander, M.~H., Manolopoulos, D.~E., Werner, H.-J., {et~al.} 2021, HIBRIDON
  is a package of programs for the time-independent quantum treatment of
  inelastic collisions and photodissociation.
\newblock \url{http://www2.chem.umd.edu/groups/alexander/hibridon}

\bibitem[{{Bacmann} {et~al.}(2012){Bacmann}, {Taquet}, {Faure}, {Kahane}, \&
  {Ceccarelli}}]{Bacmann2012}
{Bacmann}, A., {Taquet}, V., {Faure}, A., {Kahane}, C., \& {Ceccarelli}, C.
  2012, \aap, 541, L12, \dodoi{10.1051/0004-6361/201219207}

\bibitem[{{Bailey} {et~al.}(1998){Bailey}, {Chrysostomou}, {Hough}, {Gledhill},
  {McCall}, {Clark}, {Menard}, \& {Tamura}}]{Bailey1998}
{Bailey}, J., {Chrysostomou}, A., {Hough}, J.~H., {et~al.} 1998, Science, 281,
  672, \dodoi{10.1126/science.281.5377.672}

\bibitem[{Balucani {et~al.}(2018)Balucani, Skouteris, Ceccarelli, Codella,
  Falcinelli, \& Rosi}]{Balucani2018}
Balucani, N., Skouteris, D., Ceccarelli, C., {et~al.} 2018, Molecular
  Astrophysics, 13, 30

\bibitem[{{Bergantini} {et~al.}(2018){Bergantini}, {Abplanalp}, {Pokhilko},
  {Krylov}, {Shingledecker}, {Herbst}, \& {Kaiser}}]{Bergantini2018}
{Bergantini}, A., {Abplanalp}, M.~J., {Pokhilko}, P., {et~al.} 2018, \apj, 860,
  108, \dodoi{10.3847/1538-4357/aac383}

\bibitem[{{Bodo} {et~al.}(2019){Bodo}, {Bovolenta}, {Simha}, \&
  {Spezia}}]{Bodo2019}
{Bodo}, E., {Bovolenta}, G., {Simha}, C., \& {Spezia}, R. 2019, Theoretical
  Chemistry Accounts, 138, 97, \dodoi{10.1007/s00214-019-2485-3}

\bibitem[{{Corby}(2016)}]{Corby2016}
{Corby}, J. 2016, PhD thesis, University of Virginia

\bibitem[{{Corby} {et~al.}(2015){Corby}, {Jones}, {Cunningham}, {Menten},
  {Belloche}, {Schwab}, {Walsh}, {Balnozan}, {Bronfman}, {Lo}, \&
  {Remijan}}]{Corby2015}
{Corby}, J.~F., {Jones}, P.~A., {Cunningham}, M.~R., {et~al.} 2015, \mnras,
  452, 3969, \dodoi{10.1093/mnras/stv1494}

\bibitem[{{Cunningham} {et~al.}(2007){Cunningham}, {Jones}, {Godfrey}, {Cragg},
  {Bains}, {Burton}, {Calisse}, {Crighton}, {Curran}, {Davis}, {Dempsey},
  {Fulton}, {Hidas}, {Hill}, {Kedziora-Chudczer}, {Minier}, {Pracy}, {Purcell},
  {Shobbrook}, \& {Travouillon}}]{Cunningham2007}
{Cunningham}, M.~R., {Jones}, P.~A., {Godfrey}, P.~D., {et~al.} 2007, \mnras,
  376, 1201, \dodoi{10.1111/j.1365-2966.2007.11504.x}

\bibitem[{{Das} {et~al.}(2019){Das}, {Gorai}, \& {Chakrabarti}}]{Das2019}
{Das}, A., {Gorai}, P., \& {Chakrabarti}, S.~K. 2019, \aap, 628, A73,
  \dodoi{10.1051/0004-6361/201834923}

\bibitem[{{Dreiling} \& {Gay}(2014)}]{Dreiling2014}
{Dreiling}, J.~M., \& {Gay}, T.~J. 2014, \prl, 113, 118103,
  \dodoi{10.1103/PhysRevLett.113.118103}

\bibitem[{Faure {et~al.}(2019)Faure, Dagdigian, Rist, Dawes,
  Quintas-S{\'{a}}nchez, Lique, \& Hochlaf}]{Faure2019}
Faure, A., Dagdigian, P.~J., Rist, C., {et~al.} 2019, ACS Earth and Space
  Chemistry, 3, 964, \dodoi{10.1021/acsearthspacechem.9b00069}

\bibitem[{Faure {et~al.}(2013)Faure, Hily-Blant, {Le Gal}, Rist, \& {Pineau des
  For{\^{e}}ts}}]{Faure2013}
Faure, A., Hily-Blant, P., {Le Gal}, R., Rist, C., \& {Pineau des
  For{\^{e}}ts}, G. 2013, The Astrophysical Journal, 770, L2,
  \dodoi{10.1088/2041-8205/770/1/L2}

\bibitem[{{Faure} {et~al.}(2016){Faure}, {Jankowski}, {Stoecklin}, \&
  {Szalewicz}}]{Faure2016}
{Faure}, A., {Jankowski}, P., {Stoecklin}, T., \& {Szalewicz}, K. 2016,
  Scientific Reports, 6, 28449, \dodoi{10.1038/srep28449}

\bibitem[{{Faure} {et~al.}(2018){Faure}, {Lique}, \& {Remijan}}]{Faure2018}
{Faure}, A., {Lique}, F., \& {Remijan}, A.~J. 2018, The Journal of Physical
  Chemistry Letters, 9, 3199, \dodoi{10.1021/acs.jpclett.8b01431}

\bibitem[{Faure {et~al.}(2014)Faure, Remijan, Szalewicz, \&
  Wiesenfeld}]{Faure2014}
Faure, A., Remijan, A.~J., Szalewicz, K., \& Wiesenfeld, L. 2014, The
  Astrophysical Journal, 783, 72, \dodoi{10.1088/0004-637X/783/2/72}

\bibitem[{Gal(2011)}]{Gal2011}
Gal, J. 2011, Chirality, 23, 1, \dodoi{https://doi.org/10.1002/chir.20866}

\bibitem[{Garrison {et~al.}(1976)Garrison, Lester, \& Miller}]{Garrison1976}
Garrison, B.~J., Lester, W.~A., \& Miller, W.~H. 1976, The Journal of Chemical
  Physics, 65, 2193, \dodoi{10.1063/1.433375}

\bibitem[{{Glavin} \& {Dworkin}(2009)}]{Glavin2009}
{Glavin}, D.~P., \& {Dworkin}, J.~P. 2009, Proceedings of the National Academy
  of Science, 106, 5487, \dodoi{10.1073/pnas.0811618106}

\bibitem[{{Gratier} {et~al.}(2013){Gratier}, {Pety}, {Guzm{\'a}n}, {Gerin},
  {Goicoechea}, {Roueff}, \& {Faure}}]{Gratier2013}
{Gratier}, P., {Pety}, J., {Guzm{\'a}n}, V., {et~al.} 2013, \aap, 557, A101,
  \dodoi{10.1051/0004-6361/201321031}

\bibitem[{{Herbst} \& {van Dishoeck}(2009)}]{Herbst2009}
{Herbst}, E., \& {van Dishoeck}, E.~F. 2009, \araa, 47, 427,
  \dodoi{10.1146/annurev-astro-082708-101654}

\bibitem[{Herschbach \& Swalen(1958)}]{Herschbach1958}
Herschbach, D.~R., \& Swalen, J.~D. 1958, The Journal of Chemical Physics, 29,
  761, \dodoi{10.1063/1.1744588}

\bibitem[{Hollis {et~al.}(2007)Hollis, Jewell, Remijan, \& Lovas}]{Hollis2007}
Hollis, J.~M., Jewell, P.~R., Remijan, A.~J., \& Lovas, F.~J. 2007, The
  Astrophysical Journal, 660, L125, \dodoi{10.1086/518124}

\bibitem[{{Hudson} {et~al.}(2017){Hudson}, {Loeffler}, \& {Yocum}}]{Hudson2017}
{Hudson}, R.~L., {Loeffler}, M.~J., \& {Yocum}, K.~M. 2017, \apj, 835, 225,
  \dodoi{10.3847/1538-4357/835/2/225}

\bibitem[{Joyce {et~al.}(1987)Joyce, Schwartz, Miller, \& Orgel}]{Joyce1987}
Joyce, G.~F., Schwartz, A.~W., Miller, S.~L., \& Orgel, L.~E. 1987, PNAS, 84,
  4398, \dodoi{10.1073/pnas.84.13.4398}

\bibitem[{McGuire {et~al.}(2016)McGuire, Carroll, Loomis, Finneran, Jewell,
  Remijan, \& Blake}]{McGuire2016}
McGuire, B.~A., Carroll, P.~B., Loomis, R.~A., {et~al.} 2016, Science, 352,
  1449, \dodoi{10.1126/science.aae0328}

\bibitem[{{Modica} {et~al.}(2014){Modica}, {Meinert}, {de Marcellus}, {Nahon},
  {Meierhenrich}, \& {Le Sergeant d'Hendecourt}}]{Modica2014}
{Modica}, P., {Meinert}, C., {de Marcellus}, P., {et~al.} 2014, \apj, 788, 79,
  \dodoi{10.1088/0004-637X/788/1/79}

\bibitem[{M{\"{u}}ller {et~al.}(2005)M{\"{u}}ller, Schl{\"{o}}der, Stutzki, \&
  Winnewisser}]{Muller2005}
M{\"{u}}ller, H.~S., Schl{\"{o}}der, F., Stutzki, J., \& Winnewisser, G. 2005,
  Journal of Molecular Structure, 742, 215,
  \dodoi{10.1016/j.molstruc.2005.01.027}

\bibitem[{Palazzetti {et~al.}(2021)Palazzetti, Cappelletti, Coletti,
  Falcinelli, \& Pirani}]{Palazzetti21}
Palazzetti, F., Cappelletti, D., Coletti, C., Falcinelli, S., \& Pirani, F.
  2021, The Journal of Chemical Physics, 155, 234301, \dodoi{10.1063/5.0073737}

\bibitem[{Paulive {et~al.}(2021)Paulive, Shingledecker, \&
  Herbst}]{Paulive2020}
Paulive, A., Shingledecker, C.~N., \& Herbst, E. 2021, Monthly Notices of the
  Royal Astronomical Society, 500, 3414

\bibitem[{{Pizzarello} \& {Yarnes}(2018)}]{Pizzarello2018}
{Pizzarello}, S., \& {Yarnes}, C.~T. 2018, Earth and Planetary Science Letters,
  496, 198, \dodoi{10.1016/j.epsl.2018.05.026}

\bibitem[{{Remijan} {et~al.}(2008){Remijan}, {Hollis}, {Lovas}, {Stork},
  {Jewell}, \& {Meier}}]{Remijan2008}
{Remijan}, A.~J., {Hollis}, J.~M., {Lovas}, F.~J., {et~al.} 2008, \apjl, 675,
  L85, \dodoi{10.1086/533529}

\bibitem[{Requena-Torres {et~al.}(2006)Requena-Torres, Martin-Pintado,
  Rodr{\'\i}guez-Franco, Mart{\'\i}n, Rodr{\'\i}guez-Fern{\'a}ndez, \&
  De~Vicente}]{RequenaTorres2006}
Requena-Torres, M.~A., Martin-Pintado, J., Rodr{\'\i}guez-Franco, A., {et~al.}
  2006, A{\&}A, 455, 971

\bibitem[{Rivilla {et~al.}(2020)Rivilla, Mart{\'\i}n-Pintado, Jimenez-Serra,
  Mart{\'\i}n, Rodr{\'\i}guez-Almeida, Requena-Torres, Rico-Villas, Zeng, \&
  Briones}]{Rivilla2020}
Rivilla, V.~M., Mart{\'\i}n-Pintado, J., Jimenez-Serra, I., {et~al.} 2020, The
  Astrophysical Journal Letters, 899, L28

\bibitem[{{Schmiedeke} {et~al.}(2016){Schmiedeke}, {Schilke}, {M{\"o}ller},
  {S{\'a}nchez-Monge}, {Bergin}, {Comito}, {Csengeri}, {Lis}, {Molinari},
  {Qin}, \& {Rolffs}}]{Schmiedeke16}
{Schmiedeke}, A., {Schilke}, P., {M{\"o}ller}, T., {et~al.} 2016, \aap, 588,
  A143, \dodoi{10.1051/0004-6361/201527311}

\bibitem[{{\v{S}}ebest{\'{i}}k \& Bouř(2011)}]{Sebestik2011}
{\v{S}}ebest{\'{i}}k, J., \& Bouř, P. 2011, The Journal of Physical Chemistry
  Letters, 2, 498, \dodoi{10.1021/jz200108v}

\bibitem[{{Stahl} {et~al.}(2021){Stahl}, {Arenas}, {Zingsheim}, {Schnell},
  {Margul{\`e}s}, {Motiyenko}, {Fuchs}, \& {Giesen}}]{Stahl2021}
{Stahl}, P., {Arenas}, B.~E., {Zingsheim}, O., {et~al.} 2021, Journal of
  Molecular Spectroscopy, 378, 111445, \dodoi{10.1016/j.jms.2021.111445}

\bibitem[{{Thiel} {et~al.}(2017){Thiel}, {Belloche}, {Menten}, {Garrod}, \&
  {M{\"u}ller}}]{Thiel2017}
{Thiel}, V., {Belloche}, A., {Menten}, K.~M., {Garrod}, R.~T., \& {M{\"u}ller},
  H.~S.~P. 2017, \aap, 605, L6, \dodoi{10.1051/0004-6361/201731495}

\bibitem[{{Thiel} {et~al.}(2019){Thiel}, {Belloche}, {Menten}, {Giannetti},
  {Wiesemeyer}, {Winkel}, {Gratier}, {M{\"u}ller}, {Colombo}, \&
  {Garrod}}]{Thiel2019}
{Thiel}, V., {Belloche}, A., {Menten}, K.~M., {et~al.} 2019, \aap, 623, A68,
  \dodoi{10.1051/0004-6361/201834467}

\bibitem[{van~der Tak {et~al.}(2007)van~der Tak, Black, Sch{\"{o}}ier, Jansen,
  \& van Dishoeck}]{VanderTak2007}
van~der Tak, F. F.~S., Black, J.~H., Sch{\"{o}}ier, F.~L., Jansen, D.~J., \&
  van Dishoeck, E.~F. 2007, A\&A, 468, 627, \dodoi{10.1051/0004-6361:20066820}

\bibitem[{{Wang} {et~al.}(2021){Wang}, {Du}, {Semenov}, {Wang}, \&
  {Li}}]{Wang21}
{Wang}, Y., {Du}, F., {Semenov}, D., {Wang}, H., \& {Li}, J. 2021, \aap, 648,
  A72, \dodoi{10.1051/0004-6361/202140411}

\bibitem[{Wernli {et~al.}(2007a)Wernli, Wiesenfeld, Faure, \&
  Valiron}]{Wernli2007}
Wernli, M., Wiesenfeld, L., Faure, A., \& Valiron, P. 2007a, A\&A, 464, 1147,
  \dodoi{10.1051/0004-6361:20066112}

\bibitem[{{Wernli} {et~al.}(2007b){Wernli}, {Wiesenfeld}, {Faure}, \&
  {Valiron}}]{Wernli2007b}
{Wernli}, M., {Wiesenfeld}, L., {Faure}, A., \& {Valiron}, P. 2007b, \aap, 475,
  391, \dodoi{10.1051/0004-6361:20066112e}

\end{thebibliography}
\bibliographystyle{aasjournal}

\end{document}